\documentclass[preprint,aps,prd,showpacs,nofootinbib]{revtex4}
\usepackage{amsfonts}
\usepackage{mathrsfs}
\usepackage{amsmath}
\usepackage{graphicx}
\usepackage{subfigure}
\usepackage{color}

\newcommand{\bea}{\begin{eqnarray}}
\newcommand{\eea}{\end{eqnarray}}
\newcommand{\beq}{\begin{equation}}
\newcommand{\eeq}{\end{equation}}

\def\/{\over}
\def\be{\begin{equation}}
\def\ba{\begin{eqnarray}}
\def\ee{\end{equation}}
\def\ea{\end{eqnarray}}
\def\rh{r_{H}}

\begin{document}
\title{\bf Acoustic black holes in curved spacetime and emergence of analogue Minkowski metric}
\author{ Xian-Hui Ge${}^{1}$~$\footnote{ gexh@shu.edu.cn}$}
\author{ Mikio Nakahara${}^{1,2}$~$\footnote{ nakahara@shu.edu.cn}$}
\author{ Sang-Jin Sin${}^{3}$~$\footnote{ sjsin@hanyang.ac.kr}$}
\author{ Yu Tian${}^{4}$~$\footnote{ ytian@ytian@ucas.ac.cn}$}
\author{Shao-Feng Wu${}^{1}$~$\footnote{ sfwu@shu.edu.cn}$}
\affiliation{$^1$ College of Science, Shanghai University, Shanghai 200444, China\\
$^2$ Research Institute for Science and Technology, Kindai University, Higashi-Osaka, 577-8502, Japan\\
$^3$ Department of Physics, Hanyang University, Seoul  133-791, Korea\\
$^4$ School of Physics, University of Chinese Academy of Sciences, Beijing, 100049,  China
}


\vspace{3cm}

\begin{abstract}
Gravity is not only able to  be mimicked in flat spacetimes, but also in curved spacetimes.
We study analogue gravity models in curved spacetime by considering the relativistic Gross-Pitaevskii theory and Yang-Mills theory in the fixed background spacetime geometry.
The results show that acoustic metrics can  be emergent from curved spacetimes yielding a Hadamard product of a real metric-tensor and an analogue metric-tensor.
 Taking \emph{quantum vortices} as \emph{test particles}, we evaluate their released energy ratio during the ``gravitational binding". The $2+1$-dimensional flat Minkowski metric is derived from the $3+1$-dimensional Anti-de Sitter space by considering perturbations of the Yang-Mills field, which implies that Minkowski spacetime can  be also simulated and the derivations presented here have some deep connections with the holographic principle.
\end{abstract}
\maketitle

\baselineskip=16pt

Analogue models of gravity have become an active field in recent years because it provides important connections between astrophysical phenomena, such as the Hawking radiation of  black holes \cite{unruh}, the Penrose process of rotating black holes and the Kibble mechanism of topological defect deformation in the early universe, with table-top experiments. The recent experimental realizations of acoustic black hole reported was conducted in a Bose-Einstein condensate \cite{isreal,penrice,steinhauer}, optical medium \cite{ac2018} and the observation of the Kibble-Zureck mechanism was proposed in various systems \cite{kibble,zurek, chandar1994,feng2018}.

The study of analogue gravity is actually related to the exploration of the quantum nature of gravity. In 1968, Sakharov argued that gravity might be induced since \emph{it would not be fundamental from the particle physics point of view} \cite{sakharov}. Although the Weinberg-Witten no-go theorem states that  a spin-2 graviton  cannot be a composite particle in a relativistic quantum field theory \cite{WeinbergWitten},  the holographic principle and the later discovered AdS/CFT correspondence build a bridge between a $(d+1)$-dimensional gravity theory and a $d$-dimensional quantum field theory. Therefore, via the AdS/CFT correspondence, gravity can be studied in a system in the absence of gravity. The gauge/gravity duality and ¡°analogue gravity¡± are quite different but they share one
property: spacetime is emergent and gravity may not be the fundamental force. In desk-top systems, one can mimic black holes by using the fluid mechanics (i.e. acoustic black holes) in the absence of gravity. In most of the previous literature, analogue black hole metric was derived from flat Minkowski spacetime. But in general acoustic black holes can be embedded in curved spacetime. What is more, a real black hole surrounded by an acoustic horizon  is equivalent to a \emph{dumb black hole}: both sound and light cannot escape. In this paper, we are going to reverse the logic by first producing analogue gravity metric in generalized background and then try to mimic Minkowski metric from curved spacetimes.

 We first derive analogue gravity metric in curved spacetime by considering a relativistic Gross-Pitaevskii equation and then the Yang-Mills equation. The reason comes from the following observations:\\
 $\bullet$ ~~ A real black hole, a Schwarzschild black hole, for example, in the bath of the cosmological microwave background is a natural candidate for acoustic black holes.  At the location $r_s=6GM$, the escape velocity is $v_s=\frac{1}{\sqrt{3}}$ and in the region $r<r_s$, the relativistic sound waves cannot escape.\\
 $\bullet$ ~~ For astrophysical black holes, accretion disks around black holes at the centers of galaxies play a central role in explaining active galactic nuclei such as quasars and those include the most energetic steady sources of radiation in the universe. Relativistic and transonic accretion onto  astrophysical black holes is a unique example of analogue gravity realized in nature \cite{moncrief,das2004,das2004b,abraham}. \\
 $\bullet$ ~~In addition to accretion disks, black holes may be surrounded by some quantum superfluids. There has long been some proposals that dark matter might be some kinds of superfluid \cite{darkmatter}.  The transonic accretion and the condensation of the those quantum fluids also provides a scenario realizing analogue gravity.

 The structure of this paper is organized as follows: Firstly, we consider a superfluid described by the Gross-Pitaevskii equation in the curved spacetime. Secondly, we show that a 3-dimensional analogue Minkowski metric can be emerged from the combination of two $\rm SU(2)$ gauge fields in 4-dimensional Yang-Mills theory in curved spacetime. Throughout this paper, we work in the probe limit, where the background spacetime is considered as a rigid frame without dynamics. Acoustic black holes are formed only at next to leading order level. Both the Cartesian coordinates $(t,x,y,z)$ and the spherical coordinates $(t,r,\vartheta,\phi)$ will be used and switched frequently in this paper. We also adopt the units $c=G=\hbar=1$ in what follows.
\begin{figure}
\label{cone}
\includegraphics[height=4.2cm]{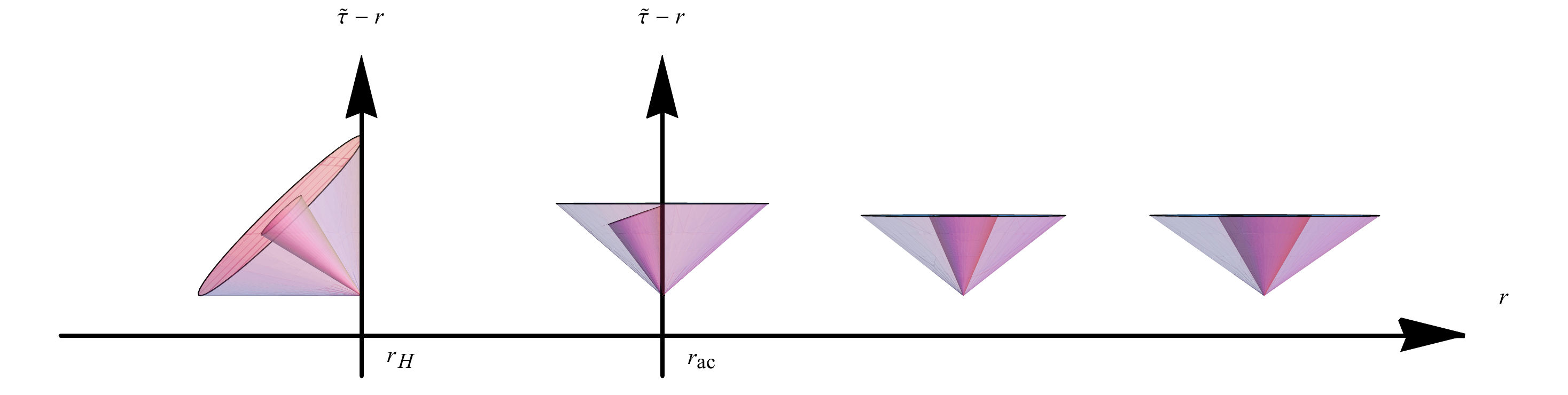}
\caption{ A cartoon of light-sound cones for an acoustic black hole in curved spacetimes. The orientation of the future light-sound cones at different radii is shown. Inside the acoustic  horizon $r_{ac}$, the sound cones becomes tilted. Further into the optical event horizon $r_{H}$, both sound and light cones become tilted.}
\vspace{-0.25cm}
\end{figure}



\section{Acoustic black holes from Gross-Pitaevskii theory in curved spacetime}

In the probe limit, the matter field has no backreaction to the background spacetime.  The Gross-Pitaevskii theory yields the action \cite{gross,GP1961}
\be\label{GP}
S=\int d^4x \sqrt{-g}\bigg(|\partial_{\mu}\phi|^2+2 m^2 |\phi|^2-b|\phi|^4\bigg),
\ee
where $\phi$ is a complex scalar order parameter.
In the followings, we  consider fluctuations of the complex scalar field.
The equation of motion for $\phi$ is written as
\be
\Box \phi+m^2\phi-b|\phi|^2\phi=0,
\ee
where  $m^2$ is a temperature dependent parameter $m^2\sim (T-T_c)$.  For temperatures above the critical temperature $T>T_c$, the phenomenological parameter $m^2$ is positive and it becomes vanishing at $T=T_c$ and negative at $T<T_c$.
The background spacetime metric is fixed as a static one
\be
ds^2=g_{tt}dt^2+g_{xx}dx^2+g_{yy}dy^2+g_{zz}dz^2.
\ee
In the Madelung representation $\phi=\sqrt{\rho(\vec{x},t)}e^{i\theta(\vec{x},t)}$, we obtain
\bea
0&=&\frac{1}{\sqrt{-g}}\partial_{t}(\sqrt{-g}g^{tt}\partial_{t}\sqrt{\rho})+\frac{1}{\sqrt{-g}}\partial_{i}(\sqrt{-g}g^{ii}\partial_{i}\sqrt{\rho})-g^{tt}\sqrt{\rho}(\partial_t \theta)^2-g^{ii}\sqrt{\rho}(\partial_i \theta)^2+m^2\rho-b\rho^{\frac{3}{2}},\nonumber\\
0&=& \frac{1}{\sqrt{-g}}\partial_{t}(\sqrt{-g}g^{tt}\rho\partial_{t}\theta)+\frac{1}{\sqrt{-g}}\partial_{i}(\sqrt{-g}g^{ii}\rho\partial_{i}\theta).
\eea
 A dumb black hole metric can be obtained by considering perturbations around the background $(\rho_0, \theta_0)$
\be
\rho=\rho_0+\rho_1,~~~ {\rm and}~~~\theta=\theta_0+\theta_1.
\ee
Working in the long-wavelength limit, thus neglecting the quantum potential terms, we obtain a relativistic wave equation governing the propagation of the phase fluctuation of the weak excitations in a homogeneous stationary condensate:
\be
\partial_{\mu}\bigg(\sqrt{-\mathcal{G}}\mathcal{G}^{\mu\nu}\partial_{\nu}\theta_1\bigg)=0,
\ee
with $\mathcal{G}=\det(\mathcal{G}_{\mu\nu})$ and the metric $\mathcal{G}_{\mu\nu}$ encodes both the information of the background spacetime metric and the background four-velocity of the fluid $v_t=-\dot{\theta}_0$, $v_{i}=\partial_i \theta_0$.
The effective metric extracted from the Klein-Gordon equation is given by
\bea \label{matrixa}\mathcal{G}_{\mu\nu}=
\mathcal{H}
\begin{pmatrix}g_{tt}(c^2_s- v^2)&\vdots&
{-v_iv_t}\cr
              \cdots\cdots \cdots\cdots&\cdot&\cdots\cdots\cdots\cdots\cdots\cdots\cr
          -v_iv_t&\vdots&{g_{ii}}(c^2_s-v_t v^t-v^2)\delta^{ij}+v_i v_j\cr
           \end{pmatrix}
, \eea
 where $\mathcal{H}=\frac{c_s}{\sqrt{c^2_s-v_{\mu}v^{\mu}}}$ and $c^2_s=\frac{b\rho_0}{2}$.

 The background four-velocity obeys the relation $b\rho_0=m^2-v_t v^t-v_i v^i$. Now we switch to the spherical coordinates $(t,r,\vartheta,\phi)$.   Under the assumption $v_i=0$, $v_t\neq 0$, $v_z=v_r$ and $g_{rr}g_{tt}=-1$, with the coordinate transformation
\be
dt=d\tilde{\tau}+\frac{v_t v_r}{g_{tt}(c^2_s-v_r v^r)}dr,
\ee we can write down the line-element for static dumb black holes \bea
ds^2&=&\frac{c_s}{\sqrt{c^2_s-v_{\mu}v^{\mu}}}\bigg[(c^2_s-v_rv^r)g_{tt}
d\tilde{\tau}^2+c^2_s \frac{c^2_s-v_{\mu}v^{\mu}}{c^2_s-v_rv^r}g_{rr}dr^2\nonumber\\&+&(c^2_s-v_{\mu}v^{\mu})g_{\vartheta\vartheta}d\vartheta^2+g_{\phi\phi}(c^2_s-v_{\mu}v^{\mu})d\phi^2\bigg],
\eea
where $\mu=0,1,2,3$. Note that we can rescale  parameters as $m^2\rightarrow \frac{m^2}{2c^2_s}$ and $v_{\mu}v^{\mu} \rightarrow \frac{v_{\mu}v^{\mu}}{2c^2_s}$, so that the metric can be recast as
\bea
ds^2&=&\frac{c^2_s}{\sqrt{1-2v_{\mu}v^{\mu}}}\bigg[(1-2v_rv^r)g_{tt}
d\tilde{\tau}^2+\frac{1-2v_{\mu}v^{\mu}}{1-2v_rv^r}g_{rr}dr^2\nonumber\\&+&(1-2v_{\mu}v^{\mu})g_{\vartheta\vartheta}d\vartheta^2+g_{\phi\phi}(1-2v_{\mu}v^{\mu})d\phi^2\bigg].
\eea
  Keeping in mind that $m^2-v_{\mu}v^{\mu}=1$ and working at the critical temperature $T=T_c$ (i.e. $m^2=0$), we can also put the metric in the form
\be \label{sonic}
ds^2=\sqrt{3}{c^2_s}\bigg[\frac{1}{3}(1-2v_rv^r)g_{tt}
d\tilde{\tau}^2+ \frac{1}{1-2v_rv^r}g_{rr}dr^2+g_{\vartheta\vartheta}d\vartheta^2+g_{\phi\phi}d\phi^2\bigg].
\ee
Thus, in general, we can write the acoustic metric in curved spacetime as a Hadamard product $\ast$  of matrices with $g^{GR}_{\mu\nu}$  denoting spacetime metric in general relativity and $g^{ac}_{\mu\nu}$ the metric of analogue gravity as
\be\label{hadmard}
ds^2=(g^{GR}\ast |g^{ac}|)_{\mu\nu}dx^{\mu}dx^{\nu},
\ee
where $||$ denotes the absolute value of $g^{ac}_{\mu\nu}$.
Although the background spacetime here is fixed to be static,  the analogue metric can be a rotating one. A Kerr-like analogue black hole was derived for relativistic fluids  in \cite{gesin,gws}.

We thus obtain a metric consists of an analogue metric multiplied by a real spacetime metric. The background real spacetime can be a Friedmann-Robertson-Walker metric or a black hole metric.
As $g^{GR}_{\mu\nu}\rightarrow \eta_{\mu\nu}$, equation (\ref{hadmard}) reduces to the acoustic black hole metric obtained in \cite{gesin}, while $g^{ac}_{\mu\nu}\rightarrow \eta_{\mu\nu}$, equation (\ref{hadmard}) becomes the spacetime  metric in general relativity. In principle, we are able to  embed analogue metric in asymptotic flat, de Sitter or Anti-de Sitter spacetimes. In the appendices, we show how to embed the analogue black hole in a ``wormhole" spacetime background. As an acoustic black hole is embedded in  real black hole spacetimes, there could be two event horizons: optical event horizon and acoustic horizon. One may notice that the metric signature has interesting observations as follows: \\
$\bullet$~For an acoustic black hole embedded in the Schwarzschild black hole background, two horizons sperate the spacetime into three regions as shown in Table \ref{table1}. It looks like that after condensation, the superfluid field makes the original black hole ``charged".\\
$\bullet$~Embedding in the Reissner-Nordstr$\ddot{o}$m geometry with outside event horizon $r_{+}$ and inner event horizon $r_{-}$ is shown in Table II. From the sign of $\mathcal{G}_{tt}$ and $\mathcal{G}_{rr}$ only, it seems that the causal structure of the original Reissner-Nordstr$\ddot{o}$m  black hole is modified. Note the fact that the acoustic metric obtained here works only in the probe limit and thus it  cannot be extended to the inside optical event horizon region. Moreover the analogue metric obtained here is not a solution of Einstein's equation.  \\
$\bullet$~Light-sound cones--
A better understanding of the metric (\ref{hadmard}) as an acoustic black hole is the behavior of sound trajectories. Sound propagates along world lines for which $d\vartheta=d\phi=0$ and $ds^2=0$.
First we introduce the tortoise coordinate $r_{*}$ defined by
\be
dr_{*}=-\frac{\sqrt{3}}{(1-2 v_r v^r)g_{tt}}dr.
\ee
Note that as $v^r\rightarrow 0$, $r_{*}$ reduces to the tortoise of optical black holes. Next, we adapt the Eddington-Finkelstein coordinate $\mathcal{V}=\tilde{\tau}+r_{*}$ and obtain
\be
ds^2=\sqrt{3}c^2_s\bigg[\frac{1}{3}(1-2v_r v^r)g_{tt}d\mathcal{V}^2+\frac{2}{\sqrt{3}}d\mathcal{V}dr\bigg].
\ee
For null geodesics $ds^2=0$, a simple solution is that some radial sound trajectories move along the curves \footnote{Here we provide a degenerated description of light rays and sound trajectories as $ds^2=0$. Actually, light rays also satisfy the null condition $ds^2=0$. }
\be
\mathcal{V}=const.~~~ ({\rm infalling})
\ee
The outgoing rays satisfy \be \label{eddin}
\frac{d\mathcal{V}}{dr}=-\frac{2\sqrt{3}}{(1-2v_r v^r)g_{tt}}.
\ee
The conditions $1-2v_r v^r=0$ and $g_{tt}=0$ characterize the location of the acoustic and optical  horizons $r_{ac}$ and $\rh$, respectively, so that sound and light rays are neither ingoing nor outgoing there, respectively (see Fig. 1 for typical sound and light cones). A special condition is that as $g_{tt}=-1$, (\ref{eddin}) only describes sound waves. \\
$\bullet$~Embedding in the Friedmann-Robertson-Walker (FRW) universe--Let us consider a complex scalar field in the early universe as given in (\ref{GP}).
 The flat FRW geometry is the simplest example of a homogeneous, isotropic cosmological metric $ds^2=-dt^2+a^2(t)(dx^2+dy^2+dz^2)$.
 We only consider the $t$-component fluctuations of the complex scalar field because at a sufficiently early time  in the inflationary epoch, the expansion rate $H$ is negligible compared to $\nabla \theta_1/a $ and the metric perturbation can be neglected \cite{wenbergbook}. In this case, the analogue metric can be expressed as\be
ds^2=[\big(2c^2_s+v^2_t\big)a^2(t)-v^2_z]\bigg\{-\frac{2c^2_s-v^2_z/a^2(t)}{\big(2c^2_s+v^2_t\big)a^2(t)-v^2_z}d\tilde{\tau}^2+\frac{2c^2_s}{2c^2_s-v^2_z/a^2(t)}dz^2+dx^2_i\bigg\}
\ee
where we have used the transformation
\be
d\tilde{\tau}=dt+\frac{v_t v_z a^2}{2c^2_s a^2-v^2_z}dz.
\ee
If we further set $v_z=0$, then we have
\be
ds^2=-2c^2_s dt^2+(2c^2_s+v^2_t)a^2(t)(dx^2+dy^2+dz^2).
\ee
 See \cite{fischer03,fischer04,fischer16,fischer17} for more references on the acoustic analog of the expanding universe produced in the Bose-Einstein condensate. 
 
 As $g_{rr}(\rh)\rightarrow  \infty$ at the optical event horizon $\rh$, the acoustic horizon
locates at $c^2_s=g_{rr}(v^r)^2$ which implies that the location of the acoustic horizon should be outside the black hole optical event horizon. Furthermore, it seems that for a black hole surrounded by an acoustic horizon, a space-like type singularity looks like a time-like singularity and vice versa.

\begin{table}\label{table1}
\centering
\begin{tabular}{@{}c|c|c@{}}
\hline
\hline
$r>r_{ac}$ &$r_{H}<r<r_{ac}$ &$r<r_{H}$ \\
\hline
$\mathcal{G}_{tt}< 0, \mathcal{G}_{rr}>0~~$  & $~~\mathcal{G}_{tt}> 0, \mathcal{G}_{rr}<0~~$ & $~~\mathcal{G}_{tt}< 0, \mathcal{G}_{rr}>0~~$ \\
\hline
$\rm both~ light~ and~ sound~ can~ escape ~~$  & $~~\rm sound~ cannot~ escape~$ & $~\rm inside~the~ black~ hole~$ \\
\hline
\hline
\end{tabular}
\caption{For an analogue metric embedded in the Schwarzschild spacetime, inside the optical event horizon $r<r_{BH}$ the sign of $\mathcal{G}_{tt}$ and $\mathcal{G}_{rr}$ are different from the neutral Schwarzschild black holes.}
\end{table}
\begin{table}\label{tabletwo}
\centering
\begin{tabular}{@{}c|c|c|c@{}}
\hline
\hline
$r>r_{ac}$ &$r_{+}<r<r_{ac}$ &$r_{-}<r<r_{+}$  &$0<r<r_{-}$\\
\hline
$\mathcal{G}_{tt}< 0, \mathcal{G}_{rr}>0~~$  & $~~\mathcal{G}_{tt}> 0, \mathcal{G}_{rr}<0~~$ & $~~\mathcal{G}_{tt}< 0, \mathcal{G}_{rr}>0~~$& $~~\mathcal{G}_{tt}> 0, \mathcal{G}_{rr}<0~~$ \\
\hline
$\rm both~light~and~sound~ can~ escape $  & $~\rm sound~ cannot~ escape$ & $~\rm inside~the~ black~ hole~$& $~\rm inside~the~ black~ hole$ \\
\hline
\hline
\end{tabular}
\caption{For an analogue metric embedded in the Reissner-Nordstr$\ddot{o}$m spacetime, inside the optical event horizon $r<r_{BH}$ the sign of $\mathcal{G}_{tt}$ and $\mathcal{G}_{rr}$ are different from the charged black holes.}
\end{table}

 The Hawking temperature of the resultant acoustic solution encodes both the information of black holes and the acoustic metric. The Hawking temperature is given by
 \be
 T=\frac{1}{4\pi\sqrt{(g^{GR}\ast g^{ac})_{rr}}}\bigg(\sqrt{\frac{g^{ac}_{tt}}{g^{GR}_{tt}}}g'^{GR}_{tt}+\sqrt{\frac{g^{GR}_{tt}}{g^{ac}_{tt}}}g'^{ac}_{tt}\bigg)_{\rm  horizon},
 \ee
 where the prime $'$ denotes derivative with respect to $r$.
 Let us consider a four dimensional black hole as an example:
$g^{GR}_{tt}=-\frac{1}{g^{GR}_{rr}}=-f(r)$ and $g^{GR}_{\vartheta\vartheta}=r^2$.  The acoustic Hawking temperature is then given by
\begin{equation}\label{Ta}
T_{a}=\frac{1}{2\sqrt{3} \pi }{ \bigg|\frac{f}{\sqrt{c_s}}(2v_rv'_rf-v^2_r
f')\bigg|}_{\rm  horizon}.
\end{equation}
We notice that at the optical event horizon where $f(\rh)=0$, the Hawking temperature of the acoustic black hole is vanishing. In general, $v_t$ cannot be zero because it is the frequency  related to relativistic dispersion relation. In the $f(r)\rightarrow 1$ limit and after recovering the scaling $v_r\rightarrow \sqrt{2} v_r/c_s$, $T_a$ can naturally
reduce to the pure acoustic black hole Hawking temperature:
$T_{a}=\frac{1}{2 \pi }{ |c'_s-v'_r|_{c_s=v_r}}$.

\subsection{Energy released during ``gravitational binding"}
It is well known that gravitational binding is a more efficient mechanism for releasing rest energy than thermonuclear mechanism. This is why black holes and compact relativistic stars are at the heart of many energetic phenomena in the universe. We are going to examine the energy released for a vortex  moving from infinity to the stable orbit of an acoustic black hole in curved spacetime. Vortices can behave as relativistic particles with their dynamics governed by the fluid metric \cite{volovik,zhang2004} and their stability ensured by a topological number. Vortices with mass given by the Einstein's relation $E=m_0 c^2_s $ \cite{popov1973,duan1994} cannot propagate at velocities faster than the sound speed $c_s$. We can therefore utilize vortices to study the time-like geodesics of massive particles near an acoustic black hole in curved spacetime.

Since the metric is independent of coordinates $t$ and $\phi$, where $\vec{u}$ is the four-velocity of the particle and $\vec{\xi}$ and $\vec{\eta}$ are Killing vectors, we can introduce energy per unit mass $\textbf{e}$ and angular momentum per unit mass $\ell$ \cite{hartle}
\bea
\textbf{e}&=&-\vec{\xi}\cdot \vec{u}=-\mathcal{G}_{tt}\frac{dt}{d\tau},\\
\ell&=&\vec{\eta}\cdot \vec{u}=\mathcal{G}_{\phi\phi}\frac{d\phi}{d\tau},
\eea
where $d\tau^2=-ds^2$ denotes the \emph{proper time}.
We can normalize the four-velocity as $\vec{u}\cdot \vec{u}=\mathcal{G}_{\alpha\beta}u^{\alpha}u^{\beta}=-1$ and this provides an integral for the geodesic equation in addition to the energy and angular momentum.
Taking account of the equatorial plane condition $u^{\theta}=0$ ($\theta=\frac{\pi}{2}$), writing $u^t=\frac{dt}{d\tau}$, $u^r=\frac{dr}{d\tau}$, $u^{\phi}=\frac{d\phi}{d\tau}$ and using the above relations to eliminate $\frac{dt}{d\tau}$ and $\frac{d\phi}{d\tau}$, we can rewrite equation (10) as
\bea
\mathcal{G}_{tt}(u^t)^2+\mathcal{G}_{rr}(u^r)^2+\mathcal{G}_{\phi\phi}(u^\phi)^2=-1.
\eea
Multiplying $\mathcal{G}_{tt}$ on both side and taking $\mathcal{G}_{\phi\phi}=r^2$, we obtain
\bea
\textbf{e}^2+\mathcal{G}_{tt}\mathcal{G}_{rr}\bigg(\frac{dr}{d\tau}\bigg)^2+\frac{\ell^2}{r^2}\mathcal{G}_{tt}=-\mathcal{G}_{tt}.
\eea
By further defining
$\mathcal{E}\equiv \frac{(e^2-1)}{2}$  and the effective potential \be
V_{\rm eff}=-\frac{1}{2}\bigg[\mathcal{G}_{tt}(1+\frac{\ell^2}{r^2})+1\bigg],
\ee
we obtain
\be\label{newton}
\mathcal{E}=-\frac{1}{2}\mathcal{G}_{tt}\mathcal{G}_{rr}\bigg(\frac{dr}{d\tau}\bigg)^2+V_{eff}.
\ee
Suppose there are some stable circular orbits outside the horizon. The angular velocity of a particle or a vortex in a circular orbit with respect to the acoustic time is the rate measured with respect to a stationary clock at infinity. For any equatorial orbit, the angular velocity is given by
\be
\Omega \equiv \frac{d\phi}{dt}=\frac{d\phi/d\tau}{dt/d\tau}=-\frac{\ell}{\textbf{e}}\frac{\mathcal{G}_{tt}}{r^2}.
\ee
The effective potential has its minimum $\partial_r V_{\rm eff}=0$ at the radius of the orbit. The value of the total energy $\mathcal{E}$ equals the value of the effective potential at the minimum.  These two requirements yields the ratio $\ell/\textbf{e}$ of circular orbits
\be
\frac{\ell}{\textbf{e}}=\bigg(-\frac{r^3 \partial_r \mathcal{G}_{tt}}{2 \mathcal{G}^2_{tt}}\bigg)^{1/2}.
\ee
The component of the four-velocity of a particle or a vortex in a circular orbit are then $u^{\alpha}=u^t(1,0,0,\Omega)$. The component $u^t$ is determined by the normalization condition $\vec{u}\cdot\vec{u}=-1$. Now there is a contribution from the angular velocity, so
\be
u^t=\left(-\mathcal{G}_{tt}+\frac{r}{2}\partial_r \mathcal{G}_{tt}\right)^{-1/2}.
\ee
The energy per unit rest mass difference between a free particle/vortex and one bound in a circular orbit of radius $r$ that is available for release is
\be
\frac{\rm released~energy}{\rm rest~energy}=1+\mathcal{G}_{tt}(-\mathcal{G}_{tt}+\frac{r}{2}\partial_r \mathcal{G}_{tt})^{-1/2}\bigg|_{r=r_{\rm ISCO}}.
\ee
For Schwarzschild metric $\mathcal{G}_{tt}=-(1-2M/r)$ with the innermost stable circular orbit (ISCO) at $r=6M$, the energy released ratio is $5.7\%$. The gravitational binding is a more efficient mechanism for releasing rest energy than thermonuclear fusion, which is approximately $1\%$. As an example, let us consider an orbit of a vortex that is the radial free fall from infinity starting from rest outside a Schwarzschild black hole. The radial component of the four-velocity $v^r$ is taken to be the escape velocity of an observer maintaining a stationary position at Schwarzschild coordinate radius $r$, that is to say $v^r \sim (\gamma\frac{2M}{r})^{1/2}$ and thus
\be
\mathcal{G}_{tt}=-\bigg(1-\frac{2M}{r}\bigg)\bigg[1- 2\gamma M/r(1-2M/r)\bigg]
 \ee by further setting  $c^2_s=\frac{1}{\sqrt{3}}$, where $\gamma$ is a tuning parameter. In this case, there are horizons with the radius $r_{H}=2M$ and $r_{ac}=(\gamma\pm\sqrt{\gamma^2-4\gamma}) M$. The acoustic horizon  $r_{ac}$ locates outside the real black hole requires $\gamma\geq 4$.
The energy released ratio are summarized in Table \ref{III}.  The energy released ratio obtained here is slightly higher than that of a Schwarzschild black hole $5.7\%$.
\begin{table}
\centering
\begin{tabular}{@{}c|c|c|c|c|c@{}}
\hline
\hline
$ $ &$~\gamma=0~$&$~\gamma=5~$ &$~\gamma=7~$ &$~\gamma=9~$ &$~\gamma=11~$\\
\hline
$~r_{\rm ISCO}~$&$~6M~$ &$~23.59 M~$ &$~35.78M~$ &$~47.86M~$ &$~59.89M~$\\
\hline
$\ell$ &$~2\sqrt{3}M~$ & $~16.70M~$ & $~23.71M~$& $~30.68M~$&$~37.63M~$ \\
\hline
$1-\textbf{e}$&$~0.057~$  & $~0.083~$ &$~0.074~$ & $~0.071~$& $~0.068~$ \\
\hline
\hline
\end{tabular}
\caption{Orbits in the equatorial plane: the radius $r_{\rm ISCO}$ of the innermost stable circular orbit (ISCO), the angular momentum per unit mass and the energy released ratio as a function of the parameter $\gamma$. M is the mass of the Schwarzschild black hole.}\label{III}
\end{table}

\section{Analogue gravity from Einstein-Yang-Mills theory}

In this section, we show how to produce a $2+1$-dimensional acoustic metric from the $3+1$-dimensional  Yang-Mills theory. To compare our results with the gauge/gravity duality theory, we demonstrate that a  $2+1$-dimensional Minkowski metric can emerge from the $\rm SU(2)$ Yang-Mills theory.

The action for the Einstein-Yang-Mills theory in $3+1$ dimensional spacetime is
\be
S=\int \sqrt{-g}d^4x\bigg[\frac{1}{2\kappa^2_4}(R+\frac{6}{l^2})-\frac{1}{2g^2_{YM}}{\rm Tr}(F_{\mu\nu}F^{\mu\nu})\bigg],
\ee
where $\kappa_4$ is the gravitational coupling, $R$ is the Ricci scalar curvature, $l$ is the radius of AdS space and $g_{YM}$ is the gauge coupling constant. The gauge field strength is given by
$F^a_{\mu\nu}=\partial_{\mu}A^a_{\nu}-\partial_{\nu}A^a_{\mu}+\epsilon^{abc}A^b_{\mu}A^c_{\nu}$ and $A=A_{\mu}dx^{\mu}=\tau^a A^a_{\mu}dx^{\mu}$. Here $\tau^a$ are the generators of $\rm SU(2)$, which obey
the relation $[\tau^a,\tau^b]=\epsilon^{abc}\tau^c$ and are related to the Pauli matrices by $\tau^a=\sigma^a/2i$. $\epsilon^{abc}$ is the totally antisymmetric tensor with $\epsilon^{123}=1$.
The Yang-Mills Lagrangian becomes ${\rm Tr}(F_{\mu\nu}F^{\mu\nu})=-F^a_{\mu\nu}F^{a\mu\nu}/2$. In the probe limit, we focus on the gauge fields only.

The equations of motion for the gauge fields are
\be
\frac{1}{\sqrt{-g}}\partial_{\mu}(\sqrt{-g}F^{a\mu\nu})+\epsilon^{abc}A^b_{\mu}F^{c\mu\nu}=0.
\ee
We consider the  symmetric metric as
\bea
&&ds^2=g_{\mu\nu}dx^{\mu}dx^{\nu}=g_{tt}dt^2+g_{xx}(dx^2+dy^2)+g_{rr}dr^2,\\
&&A=\tau^3\phi(t,x,y,r)dt+w(t,x,y,r)(\tau^1dx+\tau^2dy). \eea where
$g_{\mu\nu}$ are functions of the radial coordinate $r$ only, but
$\phi$ and $w$ are considered as functions of $(t,x,y,r)$. Such a coordinate is usually used for planar AdS black holes. The
$\rm U(1)$ subgroup of $\rm SU(2)$ generated by $\tau^3$ is identified with
the electromagnetic gauge group and $\phi$ is the electromagnetic
scalar.
 There are no analytic solutions of Einstein-Yang-Mills theory, but numerical calculation
have been widely studied (see\cite{EYM1,EYM2,volkov} for reviews).  In case
$\phi\neq 0$, but $w=0$, the solution is simply the
Reissner-Nordstr$\rm \ddot{o}$m-Anti-de Sitter (RNAdS) metric.

The (real) field $w$ is
charged under the $\rm U(1)$ group and represents the amplitudes of the
$p_x+ip_y$ components of the superfluid order parameter \cite{gubser}. In the
presence of vortices, $w$ is a complex variable and can be considered as a composite of two real fields $w=w_{R}+i w_{I}$.
We choose $A^1_{x}=A^2_y=w_{R}$ and $A^2_{x}=-A^1_{y}=w_{I}$. The $y$-component of the Yang-Mills equation for $a=1$ and $a=2$ is given by
\begin{eqnarray*}
0 & = & -\frac{1}{\sqrt{-g}}\partial_{r}(\sqrt{-g}g^{rr}g^{yy}\partial_{r}w_{I})+\frac{1}{\sqrt{-g}}\partial_{x}[\sqrt{-g}g^{xx}g^{yy}(-\partial_{x}w_{I}-\partial_{y}w_{R})]\\
 &  & +\frac{1}{\sqrt{-g}}\partial_{t}[\sqrt{-g}g^{tt}g^{yy}(-\partial_{t}w_{I}-\phi w_{R})]+g^{xx}g^{yy}w_{I}(w_{R}^{2}+w_{I}^{2})-g^{tt}g^{yy}\phi(\partial_{t}w_{R}-\phi w_{I}),\\
0 & = & \frac{1}{\sqrt{-g}}\partial_{r}(\sqrt{-g}g^{rr}g^{yy}\partial_{r}w_{R})+\frac{1}{\sqrt{-g}}\partial_{x}[\sqrt{-g}g^{xx}g^{yy}(\partial_{x}w_{R}-\partial_{y}w_{I})]\\
 &  & +\frac{1}{\sqrt{-g}}\partial_{t}[\sqrt{-g}g^{tt}g^{yy}(\partial_{t}w_{R}-\phi w_{I})]-g^{xx}g^{yy}w_{R}(w_{R}^{2}+w_{I}^{2})+g^{tt}g^{yy}\phi(-\partial_{t}w_{I}-\phi w_{R}).
\end{eqnarray*}
The $t$ component for $a=3$ is given by
\begin{eqnarray*}
\partial_{r}(\sqrt{-g}g^{rr}g^{tt}\partial_{r}\phi)+\partial_{i}(\sqrt{-g}g^{ij}g^{tt}\partial_{j}\phi)-2\sqrt{-g}g^{xx}g^{tt}[\phi(w_{R}^{2}+w_{I}^{2})+w_{R}\partial_{t}w_{I}-w_{I}\partial_{t}w_{R}]=0.
\end{eqnarray*}
In what follows, the scalar-electromagnetic potential $\phi$ is only regarded as a background field and fluctuations of $\phi$ will not be considered \cite{gesin}.
For consistency and the purpose of deriving the acoustic metric, we  assume $w_{I}$ and $w_{R}$ to be $y$-independent, so that there will be no $x$- and $y$-dependent  mixing terms. As we can see later, this will lead to an acoustic black hole metric, where the space dimension is reduced by one and this  in turn is consistent with the holographic principle, which states that the description of a volume of space can be thought of as encoded on a lower-dimensional boundary.
 Under this assumption, the equations of motion for $w_R$ and $w_I$ can be combined into a single equation
\bea
0&=&\frac{1}{\sqrt{-g}}\partial_{r}(\sqrt{-g}g^{rr}g^{yy}\partial_r w)+\frac{1}{\sqrt{-g}}\partial_{x}(\sqrt{-g}g^{xx}g^{yy}\partial_x w)+\frac{1}{\sqrt{-g}}\partial_{t}[\sqrt{-g}g^{tt}g^{yy}(\partial_t w+i\phi w)]
\nonumber\\&-&g^{xx}g^{yy}|w|^2 w+g^{tt}g^{yy}\phi(i\partial_t w-\phi w),
\eea
where $w=w_R+i w_I$.
 Notice that this is an equation in $(2+1)$-dimensional spacetime.  In the condensed phases, the
quantized field describing the microscopic system  can be replaced
by a classical mean field with a macroscopic wave function. Thus,
with the assumption $w=\sqrt{\rho(\vec{x},t)}e^{i \theta(\vec{x},t)}
$, the resulting equation of motion for the complex scalar field $w$
reduces  to real and imaginary parts, respectively \bea
\label{theta}0&=&\frac{1}{\sqrt{-g}}\partial_r\bigg(\sqrt{-g}g^{rr}g^{yy}\rho\partial_r\theta\bigg)+\frac{1}{\sqrt{-g}}\partial_{x}\bigg(\sqrt{-g}g^{xx}g^{yy}\rho\partial_x\theta\bigg)
+\frac{1}{\sqrt{-g}}\partial_t\bigg[\sqrt{-g}g^{tt}g^{yy}(\rho\partial_t\theta
\nonumber\\&+&\phi \rho)\bigg]+g^{tt}g^{yy}\frac{\partial_{t}\rho}{2},\nonumber\\
0&=&\label{rho}\frac{1}{\sqrt{-g}\sqrt{\rho}}\bigg(g^{tt}g^{yy}\partial^2_t+g^{rr}g^{yy}\partial^2_r+g^{xx}g^{yy}\frac{\partial^2_x}{2}\bigg)\sqrt{\rho}-g^{tt}g^{yy}(\partial_t\theta)^2-g^{xx}g^{yy}(\partial_x\theta)^2
\nonumber\\&&-g^{rr}g^{yy}(\partial_r\theta)^2-g^{xx}g^{yy}\rho. \eea
 When the metric $g_{\mu\nu}$
reduces to the Minkowski space $\eta_{\mu\nu}$, the above equations
are completely equivalent to those of an irrotational and inviscid
fluid apart from the quantum potential and the $\theta$-dependent
terms. In what follows, we neglect the first term in the second equation of (\ref{rho})
from the observation that the quantum potential term contains the
second derivative of slowly varying $\rho$, which is small in the
hydrodynamic region where $k,\omega$ are small.

Now linearize the fields around the background
$(\rho_0,\theta_0)$:
$\rho=\rho_0+\epsilon \rho_1$ and $\theta=\theta_0+\epsilon
\theta_1$, and write the $\mathcal{\epsilon}$ terms as
\bea
\label{ths}
0&=&\partial_{t}\bigg[\sqrt{-g}g^{tt}g^{yy}(\rho_0\partial_t\theta_1+\rho_1\partial_t\theta_0)\bigg]
+\partial_{x}\bigg[\sqrt{-g}g^{xx}g^{yy}(\rho_0\partial_x\theta_1+\rho_1\partial_x\theta_0)\bigg]
\nonumber\\&+&\partial_r\bigg[\sqrt{-g}g^{rr}g^{yy}
(\rho_0\partial_r\theta_1+\rho_1\partial_r\theta_0)\bigg],\\
0&=&2g^{tt}\partial_t\theta_0\partial\theta_1+2g^{xx}
\partial_x\theta_0\partial_x\theta_1+2g^{rr}\partial_r\theta_0\partial_r\theta_1
+\rho_1g^{xx}.\label{eight} \eea
 Defining the velocity
field by \be v_t=-\partial_t\theta_0,
~~~{v}_i=\partial_i\theta_0, \ee
where $v_t$ corresponds to the frequency obeying a ``relativistic dispersion relation": $g^{xx}\rho_0=-v_t v^t-v_r v^r-v_x v^x$£¬
we can also define \emph{three-velocity} as
\be
v^{\mu}=g^{\mu\nu}v_{\nu}.
\ee
Equations (\ref{ths}) and (\ref{eight}) can be written
as a single equation for $\theta_1$ \bea\label{com}
&&0=\partial_t\bigg\{\sqrt{-g}g^{tt}\bigg[\frac{1}{2}g^{yy}\rho_0\partial_t\theta_1+(v^{t}
\partial_t\theta_1-v^x\partial_x\theta_1-v^r
\partial_r\theta_1)v_t\bigg]\bigg\}\nonumber\\&&
+\partial_x\bigg\{\sqrt{-g}g^{xx}\bigg[\frac{1}{2}g^{yy}\rho_0\partial_x\theta_1
+\bigg(v^{t}
\partial_t\theta_1-v^x\partial_x\theta_1-v^r
\partial_r\theta_1\bigg)v_x\bigg]\bigg\}
+\nonumber\\&&\partial_r\bigg\{\sqrt{-g}
g^{rr}\bigg[\frac{1}{2}g^{yy}\rho_0\partial_r\theta_1+\bigg(v^{t}
\partial_t\theta_1-v^x\partial_x\theta_1-v^r
\partial_r\theta_1\bigg)v_r\bigg]\bigg\}
.\eea
 Comparing this with the
massless Klein-Gordon equation \be\label{KL}
\frac{1}{\sqrt{-\mathcal{G}}}\partial_{\mu}
\bigg(\sqrt{-\mathcal{G}}\mathcal{G}^{\mu\nu}\partial_{\nu}\theta_1\bigg)=0,
\ee we can extract a metric from (\ref{com}) for the sound modes\be
\sqrt{-\mathcal{G}}\mathcal{G}^{\mu\nu}\equiv
\sqrt{-g}\begin{pmatrix}g^{tt}(\frac{1}{2}g^{xx}\rho_0- v_tv^t)&\vdots&
{v^iv^t}\cr
              \cdots\cdots \cdots\cdots&\cdot&\cdots\cdots\cdots\cdots\cdots\cdots\cr
          v^iv^t&\vdots&{g^{ii}}(\frac{1}{2}g^{xx}\rho_0\delta^{ij}-v_iv^j)\cr
           \end{pmatrix}
.\label{E} \ee
By further defining the local speed of sound as \be
c^2_s=\frac{1}{2}g^{xx}\rho_0, \ee we can determine the acoustic metric simply by inverting this $3\times 3$ matrix. The acoustic metric is obtained as \bea \mathcal{G}_{\mu\nu}=
\mathcal{H}
\begin{pmatrix}g_{tt}(c^2_s- v^2)&\vdots&
{-v_iv_t}\cr
              \cdots\cdots \cdots\cdots&\cdot&\cdots\cdots\cdots\cdots\cdots\cdots\cr
          -v_iv_t&\vdots&{g_{ii}}(c^2_s-v_t v^t-v^2)\delta^{ij}+v_i v_j\cr
           \end{pmatrix}
,\nonumber \eea
 where
 \be
\mathcal{H}=\frac{(-g)^{1/4}}{\sqrt{c^2_s-(v_tv^t+v^2)}},
 \ee
 $v^2=v_i v^i$ and $g\equiv \det(g_{\mu\nu})$ is the determinant of the background spacetime metric.

   In the case $v_i=0$ and $g_{rr}=\frac{-1}{g_{tt}}$, but $v_r\neq 0$, we can simplify the metric
as \bea\label{ac} ds^2&=&(g^{GR}\ast g^{ac})_{\mu\nu}dx^{\mu}dx^{\nu}\\
&=&\mathcal{H}\bigg[(c^2_s-v_rv^r)g_{tt}
d\tau^2+c^2_s \frac{3c^2_s}{c^2_s-v_rv^r}g_{rr}dr^2+3c^2_s g_{xx}dx^2\bigg],\nonumber
\eea where the Hadamard product of matrices was used with $g^{GR}_{\mu\nu}$  representing spacetime metric in general relativity and $g^{ac}_{\mu\nu}$ the metric of analogue gravity.
Note that the coordinate transformation has been used \be
dt=d\tau+\frac{v_rv_t}{g_{tt}(c^2_s-v_rv^r)}dr. \ee

$\bullet$ \textbf{Emergent asymptotic Minkowski spacetime} The obtained metric can reduce to a pure black hole case  with an overall factor $c^2_s$, when each component of the fluid velocity vanishes (i.e. $v_t=0, v_i=0$).
The acoustic metric
in this case becomes \be
ds^2=c^2_s\bigg(g_{tt}dt^2+g_{rr}dr^2+g_{xx}dx^2\bigg). \ee
To see how the Minkowski metric is emergent, let us first assume that $\rho_0$ is  constant so that the sound velocity $c^2_s$ mainly depends on $g^{xx}$. Further setting the background spacetime to be the pure AdS space in the Poincare coordinate, one can
easily find that the resultant acoustic metric becomes
\be
ds^2={c^2_sr^2}(-dt^2+dr^2+dx^2)=\frac{\rho_0}{2}(-dt^2+dr^2+dx^2).
\ee
One observation is that one can establish a quantum field theory even in this analogue spacetime.
In case  the background spacetime has a black hole inside, the acoustic metric has its form
\be
ds^2=\frac{\rho_0}{2}\bigg(-f(r)dt^2+dx^2+\frac{1}{f(r)}dr^2\bigg),
\ee
which is also asymptotic flat as $f(r\rightarrow \infty)=1$.

$\bullet$ \textbf{FRW version}
An interesting example is that the analogue metric embedded in the flat FRW metric. The line-element of acoustic metric is listed as
\be
ds^2\sim a^{-\frac{1}{2}}(-dt^2+dx^2+dz^2).
\ee
This implies that as the universe expands the analogue metric produced by the Yang-Mills field shrinks.
\section{Conclusion and discussion}

In summary, we have shown that acoustic black holes can also be produced in curved spacetime and we use the Gross-Pitaevskii and the Yang-Mills equations for a concrete demonstration.
The analogue metric can be written as the Hadamard product of a real metric matrix and an analogue metric matrix. Working at the critical temperature can simplify the form of the metric. For acoustic black holes produced in the Gross-Pitaevskii fluid, the light-sound cone of the analogue geometry in the background of the Schwarzschild spacetime can be summarized as follows: 1). Outside the acoustic horizon, both light and sound can escape to infinity. 2). The region inside the acoustic horizon but outside the optical event horizon, light can escape although sound cannot. 3). Inside the optical event horizon of the Schwarzschild black hole, both light and sound cannot escape. Intriguingly, the optical event horizon also plays the role of acoustic horizon \footnote{Acoustic horizons naturally are not optical event horizons, but optical event horizons are naturally acoustic horizons.}. The physical picture can be understood through a thought experiment as follows: Consider a large enough black hole with small enough tidal forces so that a solid metal falling into the black hole is not to be destroyed by the tidal force. Suppose there is an observer bring a huge solid metal free falling into the acoustic black hole. For the crystal partially inside the acoustic horizon and partially outside, the sound cannot  propagate out when the observer knocks it in the acoustic horizon. But a faraway observer can still receive the light sent by the inside observer.  When the metal located  half outside and half inside the optical event horizon, knocked it again from inside, not only sound but also light cannot propagate outside the horizon. The Hawking temperature of the acoustic black hole comes from the contributions from the fluid and the background spacetime. Regarding quantum vortices as ``test particles", we calculated their energy released ratio during ``gravitational binding".

On the other hand, a $2+1$-dimensional acoustic black hole can be produced by considering the combination of two $3+1$-dimensional Yang-Mills fields. Surprisingly, this is consistent with the holographic principle. What is more, we show how the flat Minkowski metric is emergent from the bulk Anti-de Sitter space by considering perturbations of the Yang-Mills field. Considering the debates  on the fundamental nature of gravity for many years, this result indicates that both flat Minkowski spacetime and curved spacetime are not so fundamental as was considered.  The shortcoming of analogue gravity is that such models do not have a dynamical description but only reflect kinetic aspects of black hole though there are some papers trying to resolve this problem \cite{anacleto18,anacleto15,anacleto16,santos}.
It is also worth investigating analogue gravity from the holographic duality viewpoint.  In a previous paper \cite{gstz}, some of us have managed to construct an acoustic black hole from the $d$-dimensional fluid located at the timelike cutoff surface of a neutral black brane in asymptotically $\rm AdS_{d+1}$ spacetime. We also have shown that, the phonon field, which comes from the normal mode excitation of the fluid at the cutoff surface and scatters in the acoustic black hole geometry, is dual to the scalar field-the sound channel of quasinormal modes propagating in the bulk perturbed AdS black brane (see also \cite{sun17}).

\begin{acknowledgments}
 We would like to thank  Shingo Kukita and  Naoki Watamura for helpful discussions. The study was partially supported by NSFC, China ( No.11875184, No.11675097 and No.11675015); The work of M.N. is partially supported by the JSPS Grant-in-Aid for Scientific Research(KAKENHI Grant No. 17K05554); SJS was partially supported by the NRF, Korea (NRF-2016R1A2B3007687). YT is also supported by the ``Strategic Priority Research Program of the Chinese Academy of Sciences" with Grant No.XDB23030000.
\end{acknowledgments}

\begin{appendix}
\section{Acoustic non-traversable wormholes}
In \cite{unruh}, Unruh was the first to obtain an acoustic black hole metric by considering a spherically symmetric, stationary and convergent flow
\be
ds^2=\frac{\rho_0}{c_s}\bigg[-(c^2_s-v^{r2}_0)d\tau^2+\frac{c^2_s}{c^2_s-v^{r2}_0}dr^2+r^2(d\theta^2+\sin^2\theta d\phi^2)\bigg],
\ee
where $\rho_0$ is the background fluid density, $c_s$ is the local velocity of sound and $v^{r}_0$ is the radial velocity of the background fluid.
Assuming at some value of $r=R$, $v^{r}_0$ exceeds the velocity of sound $v^{r}_0=-c+\alpha (r-R)+...$, we obtain a Schwarzschild-like metric
\be
ds^2=\frac{\rho_0}{c_s}\bigg[-2\alpha c_s(r-R)d\tau^2+\frac{dr^2}{2\alpha(r-R)}+r^2(d\theta^2+\sin^2\theta d\phi^2)\bigg].
\ee
We consider a coordinate transformation $u^2=r-R$ and obtain
\be
ds^2=\frac{\rho_0}{c_s}\bigg[-2\alpha c_s u^2 d\tau^2+\frac{2 c_s du^2}{\alpha}+(u^2+R)^2(d\theta^2+\sin^2\theta d\phi^2)\bigg].
\ee
This metric is asymptotic flat as $u\rightarrow - \infty$ and $u\rightarrow \infty$, corresponding to $r\rightarrow \infty$. This is an acoustic version of the Einstein-Rosen bridge which is non-traversable for sound.
The acoustic black hole maybe useful in studying the relationship between spacetime geometry and quantum entanglement.
\section{Acoustic traversable wormholes}
It would be interesting to consider an acoustic black hole embedded in a traversable wormhole background. A 2-dimensional wormhole metric can be achieved by setting $\theta=\pi/2$ in an Ellis wormhole \cite{ellis}
\be
ds^2_{GR}=-d\tau^2+d l^2+(b^2_0+l^2)(d\theta^2+\sin^2\theta d\phi^2).
\ee
Under coordinate transformation $l^2=r^2-b^2_0$, the 2-dimensional wormhole metric is precisely equivalent to the line element of a catenoid
\be
ds^2_{GR}=-d\tau^2+\frac{r^2}{r^2-b^2_0}dr^2+r^2 d\phi^2.
\ee
An acoustic black hole embedded in a traversable wormhole is then given by
\be
ds^2=\mathcal{H}\bigg[-(c^2_s-v_rv^r)
d\tau^2+c^2_s \frac{c^2_s-v_tv^t-v_rv^r}{c^2_s-v_rv^r}\frac{r^2}{r^2-b^2_0}dr^2+(c^2_s-v_tv^t-v_rv^r)r^2d\phi^2\bigg],
\ee
where $\mathcal{H}$ is given by Eq.(\ref{matrixa}).
\end{appendix}

\end{document}